\documentclass[a4paper,11pt,headings=big,DIV=12]{article}
\pdfoutput=1 
\usepackage{graphicx}
\usepackage{dcolumn}
\usepackage{bm}
\usepackage{amssymb}
\usepackage{graphicx} 
\usepackage{amsbsy}
\usepackage{amsmath}
\usepackage{slashed}
\usepackage{amsfonts}
\usepackage{comment}
\usepackage{braket}
\usepackage{tensor}
\usepackage[scriptsize]{subfigure}
\usepackage{float}
\usepackage{enumerate}
\usepackage{multirow}
\usepackage{yfonts}

\usepackage{jheppub}

\usepackage[utf8]{inputenc}

\usepackage{listings}
\usepackage{color}

\definecolor{mygreen}{rgb}{0,0.6,0}
\definecolor{mygray}{rgb}{0.5,0.5,0.5}
\definecolor{mymauve}{rgb}{0.58,0,0.82}

\definecolor{darkWhite}{rgb}{0.94,0.94,0.94}

\newcommand{\dd} {\mathrm{d}}
\newcommand{\tr} {\mathrm{tr}}
\newcommand{\Tr} {\mathrm{Tr}}
\newcommand{\dc} {\mathcal{D}}
\newcommand{\Lagr}{\mathcal{L}}

\newcommand{\sD}{\slashed{D}}

\DeclareMathOperator{\pa}{\partial}

\newcommand{\vev}[1]{\langle #1 \rangle}

\newcommand{\al}{\alpha}
\newcommand{\be}{\beta}
\newcommand{\ga}{\gamma}

\newcommand{\de}{\delta}

\newcommand{\la}{\lambda}
\newcommand{\La}{\Lambda}
\newcommand{\eps}{\epsilon}

\newcommand{\sig}{ \sigma}
\newcommand{\Sig}{ \Sigma}

\newcommand{\EQ}{Eq.~}

\newcommand{\SEC}{Sec.~}
\newcommand{\SECs}{Secs.~}
\newcommand{\APP}{App.~}
\newcommand{\APPs}{Apps.~}

\newcommand{\TEMT}{ \Tud{\rho}{\rho} }
\newcommand{\Tud}[2]{T^{#1}_{\;\; #2}}

\newcommand{\ORD}{{\cal O}}

\newcommand{\RRt}{R \tilde R}
\newcommand{\DW}{\dc}
\newcommand{\DM}{\textfrak{D}}
\newcommand{\Log}{\log}
\newcommand{\Lam}{\La}
\newcommand{\WEYL}{trace}

\usepackage{dsfont}

\usepackage{hyperref}

\usepackage{color}
\definecolor{verdes}{cmyk}{0.92,0,0.59,0.4}

\definecolor{Grn}{rgb}{0.1,0.5,0.2}
\definecolor{Blu}{rgb}{0.,0.,0.1}
\definecolor{Red}{rgb}{0.7,0.1,0.1}
\definecolor{SE}{rgb}{0.5,0,0.4}
\definecolor{Tur}{rgb}{0,0.75,0.65}

\usepackage{hyperref} 
\hypersetup{
     colorlinks = true,
     linkcolor = black,
     citecolor = blue,
     anchorcolor = blue,
     filecolor = blue,
     urlcolor = blue,
     bookmarks=true,
     linktocpage =true,
     }

\begin{document}
{\hfill CERN-TH-2023-166}

\title{\boldmath Trace Anomaly of Weyl Fermions via the Path Integral}

\author[1]{R\'emy Larue,}
\author[1,2]{J\'er\'emie Quevillon,}
\author[3,2]{Roman Zwicky,}

\affiliation[1]{ Laboratoire de Physique Subatomique et de Cosmologie,
	 Universit\'{e} Grenoble-Alpes, CNRS/IN2P3, Grenoble INP, 38000 Grenoble, France}
\affiliation[2]{CERN, Theoretical Physics Department, Geneva 23 CH-1211, Switzerland}
\affiliation[3]{Higgs Centre for Theoretical Physics, School of Physics and
Astronomy, The University of Edinburgh, 
Peter Guthrie Tait Road, Edinburgh EH9 3FD, Scotland, UK}
\emailAdd{jeremie.quevillon@lpsc.in2p3.fr}
\emailAdd{remy.larue@lpsc.in2p3.fr}
\emailAdd{roman.zwicky@ed.ac.uk}

\abstract{
We compute the trace, diffeomorphism and Lorentz anomalies of a free Weyl fermion in a gravitational background field by path integral methods.  This is achieved by regularising the variation of the determinant of the Weyl operator building on earlier 
work by Leutwyler. 
The trace anomaly is found to be one half of the one of a Dirac fermion. Most importantly we establish that the 
 potential  parity-odd curvature term  $R \tilde R$, corresponding to the Pontryagin density, vanishes. This is to the contrary of some 
 recent findings in the literature which gave rise to a  controversy.   
 We verify,  that the regularisation does not lead to (spurious) anomalies 
 in the  Lorentz and diffeomorphism symmetries. 
 We argue that in $d = 2\;(\textrm{mod } 4)$ $P$- and $CP$-odd terms cannot appear and that for  
 $d = 4\;(\textrm{mod } 4)$ they are absent at least at leading order.
}  

\maketitle


\section{Introduction}

A classical theory free of explicit scales is invariant under scale transformation. In curved spacetime, this transformation is naturally generalised as a Weyl transformation, 
invariance implies that 
the trace of the energy-momentum tensor (EMT) vanishes, $\TEMT=0$, on 
physical states. 
At the quantum level this symmetry is anomalous as  discovered by Capper and Duff~\cite{Capper:1974ic}.\footnote{The formulation in terms of 
operators in gauge theories,  which is useful 
in other contexts,  has been worked out in Refs. \cite{Adler:1976zt,Collins:1976yq,Nielsen:1977sy}.}
Under the requirement of diffeomorphism (diffeo) invariance and on dimensional grounds, the trace anomaly  takes on the form
\begin{equation}
\label{Eq1Weyl}
g^{\al \be} \vev{T_{\al \be}}  =a\, E_4 +b\,R^2 +c\,W^2 +d\,\Box R+e\,  \RRt \;,
\end{equation}
where  $\Box$ is the Laplacian, $W^2=R_{\mu\nu\rho\sigma}R^{\mu\nu\rho\sigma}-2R_{\mu\nu}R^{\mu\nu}+\frac{1}{3}R^2$ is the Weyl tensor squared and $E_4=R_{\mu\nu\rho\sigma}R^{\mu\nu\rho\sigma}-4R_{\mu\nu}R^{\mu\nu}+R^2$ the 4-dimensional topological 
Euler density see e.g. \cite{Duff:1993wm}.
The last term is the topological  (Pontryagin) density $\RRt \equiv \frac{1}{2}\epsilon^{\mu\nu\rho\sigma}R_{\alpha\beta\mu\nu}\tensor{R}{^\alpha^\beta_\rho_\sigma}$ 
which has been  {found to be non-vanishing} in \cite{Bonora:2014qla} {and 
is known to satisfy the Wess-Zumino  consistency condition \cite{Bonora:1985cq}.} 
It is  analogous to the topological gauge field term $F\tilde F \equiv \frac{1}{2} \epsilon^{\mu\nu\rho\sigma} F_{\mu\nu} F_{\rho \sig}$ 
which appears in the ABJ anomaly \cite{Bell:1969ts,Adler:1969er,Adler:1969gk,Bardeen:1969md} 
and shares the same  $C$-, $P$- and $T$- transformation properties
\begin{equation}
\label{eq:CPT}
C \circ \RRt = +\RRt \;, \quad P \circ \RRt = -\RRt   \;, \quad T \circ \RRt = -\RRt  \;, 
\end{equation}
 such that  $\RRt$ is $P$-, $CP$-odd and $CPT$-even.  
Over the last few years a controversy has spanned in the literature since some authors 
have obtained a non-vanishing coefficient $e$  \cite{Bonora:2014qla,Bonora:2015nqa,Bonora:2017gzz,Bonora:2018ajz,Bonora:2018obr,Bonora:2022izj,Liu:2022jxz,Liu:2023yhh}  whereas others have found it to be vanishing  \cite{Bastianelli:2016nuf,Bastianelli:2019zrq,Frob:2019dgf,Abdallah:2021eii,Abdallah:2023cdw}.  Weyl fermions are subtle and probe spacetime in their own way as 
in $d = 2\;(\textrm{mod } 4)$ they give rise to gravitational (diffeomorphism and Lorentz) anomalies \cite{Alvarez-Gaume:1983ihn}, or more definitely  
to Lorentz anomalies \cite{LEUTWYLER198565,Leutwyler:1985ar}.\footnote{
The parity anomaly in QED$_{d=3}$ \cite{Redlich:1983kn} constitutes an example of an anomalous discrete symmetry. 
It is therefore not a priori clear that the Poyntryagin density could not be present.}  
In the specific determinations of $e \neq 0$ there is something similarly unsettling in that 
the authors found  it to be purely imaginary
in Lorentzian signature.
This in fact, together with \eqref{eq:CPT}, implies that its contribution is 
$CPT$-violating since $T \circ i = - i$ which would indicate a 
$CPT$ anomaly.\footnote{\label{foot:CPT} $CPT$ is a fundamental symmetry of quantum field theories and has been proven in the axiomatic context even in its curved space formulation \cite{Hollands:2009bke}.}
Whereas it was noted that an imaginary $e$ would violate unitarity  \cite{Bonora:2014qla}, the $CPT$-violation itself seems to have been overlooked.  {This would either mean that such theories have to be discarded 
 \cite{Bonora:2014qla} or supplemented by new particles such as three right-handed neutrinos in the Standard Model.}
This together with the fact that an $\RRt$-term with real coefficient in the context $P$- and or $CP$-violation is of importance for  phenomenological  reasons (see for example the discussion in \cite{Nakayama:2012gu}), e.g. baryogenesis or gravitational waves 
as reviewed in \cite{Alexander:2009tp}, and could be observed experimentally \cite{Chu:2022bhj}, we consider it important to clarify the nature of this anomaly.

In this work we show that $e=0$ when computed from the path integral by defining the Weyl determinant  building on  earlier work by Leutwyler \cite{Leutwyler:1984de,LEUTWYLER198565,Leutwyler:1985em,Leutwyler:1985ar}. The key idea is that it is the variation of the determinant, which is well-defined, that 
enters both of our approaches i) proper time regularisation and  ii) the  Fujikawa method adapted to Weyl fermions. 
In both cases this is 
  combined  with a covariant derivative expansion (CDE)
   in curved spacetime \cite{Larue:2023uyv}, which is an alternative to the heat kernel, and has already proven useful 
   to derive covariant and consistent anomalies in the context of effective field theories \cite{Filoche:2022dxl}. 
The diffeomorphism, the Lorentz and the Weyl anomalies are  entangled by regularisation, since they can be related by local counterterms (see end of \SEC\ref{sec:tech} for further details);  we therefore consider it  essential to explicitly evaluate  all three quantities.

The paper is organised as follows. In \SEC\ref{sec:tech} technicalities such as the  Weyl determinant, 
the definition of the path integral and the anomalies are discussed.  In \SECs\ref{sec:proper} 
and \ref{sec:Fujikawa} we compute anomalies applying   the proper time and  Fujikawa's method in conjunction with the CDE. 
In view of the variety of results in the literature, we make an attempt to understand them in \SEC\ref{sec:comments}. A generalisation of our results to even dimension is made in \SEC\ref{sec:d}. 
The paper ends with conclusions in \SEC\ref{sec:conc}. Technical details are deferred to \APPs \ref{App:WeylFermion} 
and \ref{App:CDEcomputations}.

\section{Technical Preliminaries}
\label{sec:tech}

\subsection{The determinant of the Weyl operator}
\label{sec:detD}

There are several challenges in defining the determinant for a Weyl fermion. 
For example if one starts with a  Dirac fermion with only left-handed components 
then its associated Dirac operator $i\slashed{D} P_L$, where $P_L$ projects on left-handed fermions, 
 cannot be inverted. 
This problem can be avoided if one starts directly with a two-component Weyl fermion $\psi_L$.
The Weyl operator $\DW$, which is the Dirac operator acting on a Weyl fermion, is given by
\begin{equation}
\label{eq:DW}
\DW \psi_L = i \bar\sig^\mu  (\partial + \omega_L
)_\mu \psi_L  \;,
\end{equation}
where $\bar\sig^\mu = \tensor{e}{^\mu_a}\bar \sig^a$, 
$\bar\sig^a = (1, -\vec{\sig})$,  $ \tensor{e}{^\mu_a}$ is the vierbein   
 and  $\omega_L$ is the spin-connection \eqref{eq:DLR}; more precise 
 definitions can be found in the \APP\ref{App:WeylFermion}.
Hereafter we work in  Euclidian space as it is technically more convenient. 
The determinant of the Weyl operator appears formally in the effective action 
\begin{equation}
\label{eq:W}
W=- \Log \det \DW \;,
\end{equation}
after performing the Gaussian path integral.
 Unfortunately, 
$\det \dc$ is ill-defined, as emphasised by 
\'Alvarez-Gaum\'e and Witten \cite{Alvarez-Gaume:1983ihn}, since it maps left onto right handed fermions which have different Hilbert space, i.e. $\DW:  (\frac{1}{2},0) \to (0,\frac{1}{2})$ and vice versa. This makes the phase of the functional determinant ambiguous, whereas
the modulus is unaffected (and is also gauge invariant). However,  the determinant itself is not an observable.
That is  Leutwyler and Mallik   \cite{Leutwyler:1984de,LEUTWYLER198565,Leutwyler:1985em,Leutwyler:1985ar}  pointed out that the  variation of  \eqref{eq:W}
\begin{equation}
\label{eq:Wformal}
\delta \Log\det {\DW}=\Tr\,\de \DW \DW^{-1}\;,
\end{equation}
which formally holds for any operator, is well-defined since it maps fermions to fermions of the same chirality: 
$\de \DW \DW^{-1}:  (0,\frac{1}{2}) \to (0,\frac{1}{2}) $.  
This is in line with the observation that 
the relative phase between two operators is well-defined \cite{Alvarez-Gaume:1983ihn}.  
{In both the proper time regularisation and the Fujikawa method adapted for Weyl fermion it will be 
the formula \eqref{eq:Wformal} and not the determinant itself which will form the starting point of the evaluation.}

{The zero modes of $\DW$ consist in another problem for the definition of the determinant.  
In what follows we will assume that there are no zero modes, as is often done \cite{Leutwyler:1985em,Witten:2019bou}, which is believed to be true in the realm of perturbation theory.}
On top of this the operator $\DW^{-1}$ is singular, as apparent from perturbation theory at short distances, but can be regularised to which we will turn to in \SECs\ref{sec:proper} and 
\ref{sec:Fujikawa} respectively.

\subsection{Path integral formulation}
\label{sec:PI}

For the remaining part of the paper, Fujikawa's diffeo-invariant path integral measure 
\cite{Fujikawa:1980rc,Fujikawa:2004cx}  is used.  
It  is free from spurious gravitational anomalies (i.e that can be removed by local counterterms) in any even dimension. It has been shown for Dirac fermions \cite{Fujikawa:1980rc,Fujikawa:2004cx}
and in this paper we show that it equally holds for Weyl fermions.

Since we will first compute the anomaly for a Dirac fermion\footnote{In the absence of charges a Dirac fermion is equivalent to what is known as a vector-like fermion.}, validating the method, 
we  have to define the Dirac operator, in analogy to 
Weyl fermions  in \eqref{eq:DW}. It reads
\begin{equation}
\dc\psi =i\sD\psi= i\gamma^\mu(\pa+\omega
)_\mu\psi\; ,
\end{equation}
where the spin-connection is $\omega_\mu=\frac{1}{2}\omega_{\mu,ab}\Sigma^{ab}$, with $\omega_{\mu,ab}$ the tangent frame spin-connection and $\Sigma^{ab}=\frac{1}{4}[\gamma^a,\gamma^b]$. Latin indices denote the tangent space indices, whereas greek indices are the base manifold indices. The Dirac matrices in curved spacetime follow from the inverse vierbein $\gamma^\mu=\tensor{e}{^\mu_a}\gamma^a$. The covariant derivative is compatible with the vierbein, the vierbein determinant and the Dirac matrices: $D_\mu\tensor{e}{^\nu_a}=D_\mu\,e=D_\mu\gamma^\nu=0$. It  acts on tensors (with no spinorial indices) as
\begin{equation}
\label{eq:Christoffel}
D_\mu t^\nu=\pa_\mu t^\nu+\Gamma^\nu_{\mu\rho}t^\rho\;, \quad\quad D_\mu t^a=\pa_\mu t^a +\tensor{\omega}{_\mu^a_b}t^b \;,
\end{equation}
 and we follow \cite{Misner:1973prb} for the conventions in gravity.

Fujikawa's rescaled fermionic variables  are defined by ($e=\det \tensor{e}{^a_\mu}$)
\begin{equation}
\tilde\psi=\sqrt{e}\psi \;, \quad \tilde{\bar\psi}=\sqrt{e}\bar\psi \;, 
\end{equation}
for  which the measure $\DM\tilde{\bar\psi}\DM\tilde\psi$ is  diffeo-invariant in $d = 4\;(\textrm{mod } 4)$.
The Dirac operator associated with the adjusted spinors is
\begin{equation}
\label{Dpsitilde}
\dc\tilde\psi=i\gamma^\mu\left(\pa_\mu+\omega_\mu-\frac{\pa_\mu\sqrt{e}}{\sqrt{e}}\right)\tilde\psi \;.
\end{equation}
Similarly, the Weyl fermion invariant measure is $\DM\tilde{\bar\psi}_L\DM\tilde\psi_L$, and the Weyl operator reads
\begin{equation}\label{eq:DW2}
\dc\tilde\psi_L=i\bar\sigma^\mu\left(\pa_\mu+\omega^L_\mu-\frac{\pa_\mu\sqrt{e}}{\sqrt{e}}\right)\tilde\psi_L \;.
\end{equation}
The corresponding path integrals assume the form
\begin{equation}
\label{eq:Fujikawa}
Z_{\textrm{Dirac}} =\int \DM \tilde{\bar\psi} \DM \tilde\psi e^{-\int\dd^4x \tilde{\bar\psi} \dc \tilde \psi}\; ,\quad 
Z_{\textrm{Weyl}} =\int \DM \tilde{\bar\psi}_L \DM \tilde\psi_L e^{-\int\dd^4x \tilde{\bar\psi}_L \dc \tilde \psi_L}\; ,
\end{equation}
where the $\tilde \psi$- and $\tilde \psi_L$-variables are to be treated as vierbein-independent.

\subsection{Definition of the anomalies}
\label{sec:anomal}

Finally, we turn to the definition of the anomalies  which we will also express in terms of the EMT.  
It is well-known that the anomalies follow from a variation of the effective quantum action \eqref{eq:W} with 
respect to a symmetry. Concretely, applying $\de_{\al(x)}$ to \eqref{eq:W}, using \eqref{eq:Wformal}, the 
associated anomaly $\mathcal{A}$ is formally defined by
\begin{equation}
\de_{\al}  W = -\Tr\,\de_{\al} \DW\DW^{-1}\, = -
\int\dd^4x\,e\,\alpha(x)\mathcal{A}  \;.
\label{LeutwylerA}
\end{equation}
We define the three symmetry transformations, Weyl, diffeo and Lorentz,  on the field variables.\footnote{We describe these transformations on the fields that occur in the Dirac case, from which we will infer  the transformations involving a Weyl fermion and its Weyl operator later on.} 
The Weyl variation of infinitesimal parameter $\sig(x)$   reads
\begin{alignat}{6}
&\delta^{W}_\sigma\tensor{e}{^\mu_a} &\;=\;& -\sigma\, \tensor{e}{^\mu_a}\;, \quad   & & \delta^{W}_\sigma e&\;=\;& d\,\sigma\, e
\;, \quad   & & \delta^{W}_\sigma \omega_\mu&\;=\;& \frac{d-1}{2}(\pa_\mu\sigma) \;,
 \nonumber \\[0.1cm]
&\delta^{W}_\sigma\tilde\psi &\;=\;& \frac{1}{2}\sigma\,\tilde\psi \;, \quad   & & \delta^{W}_\sigma\tilde{\bar\psi} &\;=\;&
\frac{1}{2}\sigma\,\tilde{\bar\psi} \;. & & 
\end{alignat}
The (active) diffeo-transformation of infinitesimal parameter $\xi_\mu(x)$ reads
\begin{alignat}{4}
&\delta^{d}_\xi\tilde\psi&\;=\;&\xi^\mu\pa_\mu\tilde\psi+\frac{1}{2}(\pa_\mu\xi^\mu)\tilde\psi
\;, \quad   & &  \delta^{d}_\xi\tilde{\bar\psi}&\;=\;&\xi^\mu\pa_\mu\tilde{\bar\psi}+\frac{1}{2}(\pa_\mu\xi^\mu)\tilde{\bar\psi} \;, \nonumber \\[0.1cm]
&\delta^{d}_\xi\tensor{e}{^\mu_a}&\;=\;&\xi^\nu\pa_\nu\tensor{e}{^\mu_a}-\tensor{e}{^\nu_a}\pa_\nu\xi^\mu\;, \quad   & &
\delta^{d}_\xi e&\;=\;&(\pa_\mu e\xi^\mu) = e (D_\mu\xi^\mu)  \;, \nonumber \\[0.1cm]
&\delta^{d}_\xi\omega_\mu&\;=\;&\xi^\nu\pa_\nu\omega_\mu+\omega_\nu\pa_\mu\xi^\nu \;.
\end{alignat}
The Lorentz transformation of infinitesimal parameter $\alpha_{ab}(x)$ ($\alpha_{ab}=-\alpha_{ba}$) reads
\begin{alignat}{4}
&\delta^{L}_\alpha\tilde\psi &\;=\;& -\frac{1}{2}\alpha_{ab}\Sigma^{ab}\, \tilde\psi 
\;,\quad \delta^{L}_\alpha\tilde{\bar\psi}=\frac{1}{2}\alpha_{ab}\,\tilde{\bar\psi} \Sigma^{ab}
\;, \quad \delta^{L}_\alpha\tensor{e}{^\mu_a}=\tensor{e}{^\mu_b}\tensor{\alpha}{^b_a}
\;, \quad   & & \delta^{L}_\alpha e&\;=\;& 0 \;, \nonumber \\[0.1cm]
&\delta^{L}_\alpha \omega_\mu &\;=\;& \frac{1}{2}[D_\mu,\alpha_{ab}\Sig^{ab}]=\frac{1}{2}\Sig^{ab}[D_\mu,\alpha_{ab}] \;.
\end{alignat}
As stated above we further define the anomalies from the path integral
\begin{equation}
\delta_\al W =-\frac{\int\dc\tilde\psi\dc\tilde{\bar\psi} \left(\delta_\al S\right) e^{-S}}{\int\dc\tilde\psi\dc\tilde{\bar\psi} e^{-S}} 
=-\langle \delta_\al S\rangle  \;,
\end{equation}
where $S$ is the action. 
Since $W$ is a functional of the background fields only,  $\delta_\al$ acts solely on the background fields, which 
in the case at hand are the vierbeins.
For the Weyl, the diffeo and the Lorentz transformations of respective parameters $\sig$, $\xi^\mu$ and $\al_{ab}$ one obtains 
\begin{alignat}{3}
&  \int\dd^4x\,e\,\sigma\, \mathcal{A}_{\mathrm{\WEYL}}  &\;=\;& - \delta^{W}_\sigma W  &\;=\;&   \int\dd^4x e\, \sig \, \tensor{e}{^a_\mu}\langle\tensor{T}{^\mu_a} \rangle  
\;, \nonumber \\[0.1cm]
& \int\dd^4x\,e\,\xi_\mu\, \mathcal{A}_\mathrm{diffeo}^{\mu} &\;=\;&- \delta^{d}_\xi W   &\;=\;&    \int\dd^4x e \, \xi^\nu \left \langle  \tensor{\omega}{_\nu^a^b}T_{ab} - 
D^\mu T_{\mu\nu}  \right\rangle 
  \;, \nonumber \\[0.1cm]
&\int\dd^4x\,e\,\alpha_{ab} \, \mathcal{A}^{ab}_{\mathrm{Lorentz}}  &\;=\;&-\delta^L_\alpha W  &\;=\;&  \int\dd^4x e \, \alpha_{ab} \langle T^{ab} \rangle  \;,
\end{alignat}
where the EMT is obtained from 
$e \, \tensor{T}{^\mu_a}=\delta S/\delta \tensor{e}{^a_\mu}$, 
treating $\tilde\psi$, $\tilde{\bar\psi}$ as vierbein-independent.   
The tensors $T_{\mu\nu}$ and  $T_{ab}$ follow from $\tensor{T}{^\mu_a}$ using the vierbein and  the metric accordingly.
Note that 
the EMT is not automatically symmetric, as it  is when  varying the action by the metric, 
since with fermions it is the vierbein  that becomes the fundamental quantity. 
This allows for the Lorentz anomaly to be present in the first place  and  as a result the diffeo anomaly  is not given solely by the divergence of the EMT~\cite{Bertlmann:1996xk,Fujikawa:2004cx}.

It is known that the Lorentz and diffeo anomalies are related by a local counterterm \cite{Alvarez-Gaume:1983ict,Bardeen:1984pm}. However, as pointed out in \cite{LEUTWYLER198565,Leutwyler:1985ar}, this local counterterm is non-polynomial in the background fields and not permitted by the rules of renormalisation, which means that a priori one can not always transfer the Lorentz anomaly into the diffeo one and conversely.  
In principle it is known that in $d = 4\;(\textrm{mod } 4)$, but not $d = 2\;(\textrm{mod } 4)$, these anomalies 
are vanishing \cite{Alvarez-Gaume:1983ihn,LEUTWYLER198565,Leutwyler:1985ar}. 
In fact in $d=2$ Leutwyler \cite{LEUTWYLER198565} has shown that for a Weyl fermion, there is only a Lorentz anomaly and no diffeo anomaly, and it is therefore likely that this pattern repeats for $d = 2\;(\textrm{mod } 4)$.
However, in view of the controversy around the $\RRt$-term we verify  the absence of spurious Lorentz and diffeo anomalies explicitly.

\section{Proper Time Regularisation}
\label{sec:proper} 

In order to perform a concrete computation  the singular operator 
$\DW^{-1}$, in \eqref{eq:Wformal}, has to be regularised. Following \cite{Leutwyler:1985ar} we use the proper time regularisation\footnote{Alternatively, $\dc^{-1}|_{\Lam}=\int_{\Lam^{-2}}^\infty \dd t\, e^{-t \dc^\dagger\dc}\dc^\dagger$ could have been chosen 
} 
\begin{equation}
\label{eq:proper}
\DW^{-1}|_{\Lam}=\int_{\Lam^{-2}}^\infty \dd t \, \DW^\dagger e^{-t \DW\DW^\dagger}\;,
\end{equation}
which is convergent in the infrared, that is $t \to \infty$, since  $ \DW\DW^\dagger$ is a manifestly positive operator, but still requires an ultraviolet regulator $\Lam$.
Inserting this expression above, the integral easily evaluates to 
\begin{equation}
\label{eq:lnreg}
\delta\log\det\DW|_{\Lam} = \Tr\,\delta\DW \int_{\Lam^{-2}}^\infty \dd t \,\DW^\dagger e^{- t \DW\DW^\dagger}=\Tr\,\delta\DW\DW^{-1}e^{-\frac{\DW\DW^\dagger}{\Lam^2}}\; .
\end{equation}
It is well-known that the anomalies are exactly marginal, that is  $\Lam^0$-terms. 
The divergences in $\Lam^4$ and $\Lam^2$ are
 of no special interest  
and will therefore  not be discussed any further.
In what follows we will apply this regularisation to a Dirac and a Weyl fermion. The Dirac case is beyond doubt in the literature but serves to test 
and illustrate the method. 

\subsection{Dirac fermion }
\label{sec:vector}

For a Dirac fermion the operator $\DW$ assumes the form 
\begin{equation}
\DW \to i\tensor{e}{^\mu_a}\gamma^a D_\mu = i \slashed{D} \;,
\end{equation}
where   $D_\mu$ is given by \eqref{Dpsitilde}, when acting on $\tilde\psi$.
The fact that it is hermitian $\DW^\dagger=\DW$, makes this case 
 particularly simple since the regulator  $e^{- t \DW\DW^\dagger} \to e^{- t \DW^2} $  then commutes with both $\DW$ and $\DW^{-1}$.
The Weyl variation reads
\begin{equation}
\delta^W_\sigma\DW=-\sigma\DW-\frac{1}{2}[\DW,\sigma]\;.
\end{equation}
Therefore the trace anomaly is given by
\begin{alignat}{2}
& \delta^W_\sig W &\;=\;& \lim_{\Lam \to \infty}\Tr\,\left(\sigma\DW+\frac{1}{2}[\DW,\sigma]\right)\DW^{-1}e^{-\frac{\DW^2}{\Lam^2}} \\[0.1cm]
& &\;=\;& \lim_{\Lam \to \infty}\Tr\,\left(\sigma e^{-\frac{\DW^2}{\Lam^2}}-\frac{1}{2}\sigma[\DW,\DW^{-1}e^{-\frac{\DW^2}{\Lam^2}}]\right) 
= \lim_{\Lam \to \infty}\Tr\,\sigma e^{-\frac{(i\sD)^2}{\Lam^2}}\;,\nonumber
\end{alignat}
where the cyclicity of the trace has been used. 
  To evaluate the last term we use 
the CDE  and obtain
\begin{equation}
\mathcal{A}_{\mathrm{\WEYL}}^{\mathrm{Dirac}} = \frac{1}{16\pi^2}\bigg(\frac{1}{72}R^2-\frac{1}{45}R_{\mu\nu}R^{\mu\nu}-\frac{7}{360}R_{\mu\nu\rho\sigma}R^{\mu\nu\rho\sigma} 
-\frac{1}{30}\Box R 
\bigg)\;,
\end{equation}
which is the well-known result \cite{Bertlmann:1996xk,Fujikawa:2004cx} first obtained in \cite{Capper:1974ic}.
It remains to verify that there are no spurious Lorentz and diffeo anomalies.
Using $\delta^L_\alpha \DW=\frac{1}{2}[\DW,\alpha_{ab}\Sig^{ab}]$, the Lorentz-variation reads
\begin{equation}
\delta^L_\alpha W |_{\Lam}=  -\Tr\,\delta^L_\alpha\DW \DW^{-1}e^{-\frac{\DW^2}{\Lam^2}} = \Tr\,\frac{1}{2}\alpha_{ab}\Sig^{ab}[\DW,\DW^{-1}
e^{-\frac{\DW^2}{\Lam^2}} ]=0 \;,
\end{equation}
and thus $ \mathcal{A}^{ab}_{\mathrm{Lorentz}}=0$ follows. 
The diffeo-variation of $\DW$ reads
\begin{equation}
\delta^d_\xi\DW=-[\DW,\xi^\mu]\nabla_\mu-\xi^\mu[\DW,\nabla_\mu]-\frac{1}{2}[\DW,(D_\mu\xi^\mu)]\;,
\end{equation}
where $\nabla$ is the covariant derivative deprived of 
spin-connection\footnote{$\nabla$ only contracts indices in the tangent and base spaces as in \eqref{eq:Christoffel}. In particular we have $D_\mu\xi^\nu=\nabla_\mu\xi^\nu$.} 
\begin{equation}
\nabla_\mu=D_\mu-\omega_\mu
 \;.
\end{equation}
The diffeo-variation reads
\begin{alignat}{2}
&\delta^d_\xi W|_{\Lam} &\;=\;&- \Tr\,\delta^d_\xi\DW\DW^{-1}e^{-\frac{\DW^2}{\Lam^2}}  \\[0.1cm]
& &\;=\;& -\Tr\,\bigg\{\xi^\mu[\DW,\nabla_\mu \DW^{-1}e^{-\frac{\DW^2}{\Lam^2}}]-\xi^\mu[\DW,\nabla_\mu]\DW^{-1}e^{-\frac{\DW^2}{\Lam^2}} +\frac{1}{2}(\nabla_\mu\xi^\mu)[\DW,\DW^{-1}e^{-\frac{\DW^2}{\Lam^2}}]\bigg\} \nonumber \\[0.1cm]
& &\;=\;&  -\Tr\,\bigg\{\xi^\mu \DW \nabla_\mu\DW^{-1}e^{-\frac{\DW^2}{\Lam^2}}-\xi^\mu \nabla_\mu e^{-\frac{\DW^2}{\Lam^2}} -\xi^\mu\DW \nabla_\mu\DW^{-1}e^{-\frac{\DW^2}{\Lam^2}}+\xi^\mu \nabla_\mu e^{-\frac{\DW^2}{\Lam^2}}\bigg\}=0 \;. \nonumber
\end{alignat}
It is noted that both the Lorentz and diffeo anomalies are vanishing prior to  taking the limit $\Lam \to \infty$.
That the Lorentz and diffeo symmetry are not anomalous 
for Dirac fermions in any even dimension is a known result and further validates the method. 
For example, had we used  the standard path integral measure $\DM\psi\DM\bar\psi$, 
instead of Fujikawa's  \eqref{eq:Fujikawa}, 
{we would have obtained a non-vanishing but spurious diffeo anomaly and the wrong trace anomaly \cite{Fujikawa:2004cx}.
The correct one would then be obtained by adding a local counterterm that removes the spurious diffeo anomaly 
and leads to the correct trace anomaly.  Note that no spurious Lorentz anomaly would arise since
the determinant of the vierbein is Lorentz-invariant.}

\subsection{Weyl fermion}
\label{sec:Weyl}

In this section and the remaining part of the paper, $\dc$ is the Weyl operator and is given by \eqref{eq:DW2}. In order to compute the variation of the Weyl operator $\de \DW$ it is easier to evaluate  its Dirac counterpart $ \de i\sD$  as in the previous section and then project it in the left-right basis using Eq. \eqref{eq:iDtoLR}. We obtain
\begin{alignat}{2}
\label{eq:var}
&\delta^W_\sigma\DW &\;=\;& -\sigma \DW-\frac{1}{2}[\DW,\sigma]\;,\\[0.1cm]
&\delta^d_\xi\DW &\;=\;& -[\DW,\xi^\mu]\nabla_\mu-\xi^\mu [\DW,\nabla_\mu]-\frac{1}{2}[\DW,(D_\mu\xi^\mu)]\nonumber\;,\\[0.1cm]
&\delta^L_\alpha \DW &\;=\;& [\DW ,\frac{1}{2}\alpha_{ab}\mu^{ab}]+\frac{1}{2}\alpha_{ab}(\mu^{ab}-\lambda^{ab})\DW \;, \nonumber
\end{alignat}
where $\mu^{ab}$ and $\lambda^{ab}$ are defined in App.\ref{App:WeylFermion},   and we note that their form is equivalent to the Dirac case except for the Lorentz transformation. 

Let us first verify that the diffeo and Lorentz anomalies vanish, such that the regularisation induces no spurious gravitational anomaly. The diffeo anomaly is given by
\begin{equation}
\label{WeylFermiondeltacov1}
\delta^{d}_\xi W \;=\;-\lim_{\Lambda\to\infty}\Tr\,\delta^d_\xi\DW\, \DW^{-1}\,e^{-\frac{\DW\DW^\dagger}{\Lambda^2}}\\
\;=\;-\lim_{\Lambda\to\infty}\Tr\,\left(\xi^\mu\nabla_\mu + \frac{1}{2}(D_\mu\xi^\mu)\right)\left(e^{-\frac{\DW\DW^\dagger}{\Lambda^2}}-e^{-\frac{\DW^\dagger\DW}{\Lambda^2}}\right)\;,
\end{equation}
where the cyclicity of the trace, \EQ\eqref{eq:DRsigma} and
\begin{equation}
\label{eq:crucial}
\DW^{-1}e^{-\frac{\DW\DW^\dagger}{\Lambda^2}}\DW=e^{-\frac{\DW^\dagger\DW}{\Lambda^2}}\;,
\end{equation}
have been used.
Noting that
\begin{align}
&i\sD=\begin{pmatrix}
0 & \DW \\
\DW^\dagger & 0 \\
\end{pmatrix},\qquad\qquad e^{-\frac{ (i\sD)^2}{\Lam^2}}=\begin{pmatrix}
e^{-\frac{\DW\DW^\dagger}{\Lambda^2}} & 0 \\
0 & e^{-\frac{\DW^\dagger\DW}{\Lambda^2}} \\
\end{pmatrix}\;,
\label{WeylToDirac}
\end{align}
we can recast \EQ\eqref{WeylFermiondeltacov1} in Dirac space
\begin{align}
\label{deltad_Weyl}
\delta^d_\xi W=-\lim_{\Lambda\to\infty}\Tr\left(  \gamma_5\left(\xi^\mu \nabla_\mu+\frac{1}{2}(D_\mu\xi^\mu)\right)e^{-\frac{ (i\sD)^2}{\Lam^2}}\right)\;.
\end{align} 
A direct computation using the CDE in curved spacetime shows that it vanishes.
Importantly, the diffeo anomaly may a priori not be covariant since $\nabla$ in \eqref{deltad_Weyl} is not the covariant
derivative. 
This is where the CDE as carried out in \cite{Larue:2023uyv} 
is useful since it easily allows for an expansion that is not manifestly covariant; 
with more details   in \APP\ref{App:CDEcomputations}.
{It is noted that the heat kernel with Seeley-DeWitt coefficients  \cite{Schwinger:1951nm,DeWitt:1967yk,DeWitt:1967ub}  
and former CDE in curved spacetime approaches \cite{Binetruy:1988nx,Alonso:2019mok} are designed to compute traces involving a quadratic operator, such as $\Tr\, a(x) e^{-D^2}$ where $a$ is not a differential operator. With some work a trace of the form \eqref{deltad_Weyl} can be brought to this form, but it is not straightforward and involves lengthy manipulations \cite{Leutwyler:1985ar}.

The Lorentz anomaly is given by
\begin{align}
\begin{split}
\delta^L_\alpha W=-\lim_{\Lambda\to\infty}\Tr\,\frac{1}{2}\alpha_{ab}\left(\mu^{ab}e^{-\frac{\DW^\dagger\DW}{\Lambda^2}}-\lambda^{ab}e^{-\frac{\DW\DW^\dagger}{\Lambda^2}}\right)\;,
\end{split}
\end{align}
and once again it can be rewritten as a trace in Dirac space
\begin{align}
\begin{split}
\label{deltaL_Weyl}
\delta^L_\alpha W=\lim_{\Lambda\to\infty}\Tr\,\frac{1}{2}\alpha_{ab}\Sigma^{ab}\gamma_5e^{-\frac{ (i\sD)^2}{\Lam^2} }\;.
\end{split}
\end{align}
The direct computation using the CDE shows that it equally vanishes.

We now turn to the trace anomaly. Using \eqref{eq:crucial} one obtains 
\begin{align}
\label{eq:WW}
\delta^W_\sigma W=\frac{1}{2}\lim_{\Lambda\to\infty}\Tr\,\sigma \left(e^{-\frac{\DW^\dagger\DW}{\Lambda^2}}+e^{-\frac{\DW\DW^\dagger}{\Lambda^2}}\right)=\frac{1}{2}\lim_{\Lambda\to\infty}\Tr\,\sig e^{-\frac{ (i\sD)^2}{\Lam^2}} \; ,
\end{align}
which is half the trace anomaly of a Dirac fermion
\begin{equation}
\mathcal{A}^{\mathrm{Weyl}}_{\mathrm{trace}}=\frac{1}{2}\mathcal{A}^{\mathrm{Dirac}}_{\mathrm{trace}}
\;,
\end{equation}
without spurious gravitational anomalies,
\begin{equation} 
\mathcal{A}_{\mathrm{diffeo}}^\mu=\mathcal{A}_{\mathrm{Lorentz}}^{ab}=0\;.
\end{equation}
In particular there is no Pontryagin density $R\tilde R$. We wish to emphasise that  each term in \eqref{eq:WW}  
has an $\RRt$-component, but it cancels in the sum of the two. This is in agreement with the computation of the heat kernel coefficient $b_4$ of the representation $(1/2,0)$ of the Lorentz group \cite{Christensen:1978md}, and we showed that the trace anomaly of a Weyl fermion is determined by $b_4(1/2,0)+b_4(0,1/2)$.
We further note that the second term in \eqref{eq:WW} 
originates from the second one in \EQ\eqref{eq:var} which in turn is due to the spin-connection.

We note that this result has previously been obtained in a similar setting by Leutwyler and Mallik \cite{Leutwyler:1985ar}. 
The  main difference is in the evaluation of the expression in \eqref{deltad_Weyl} for which they use 
the heat kernel which is rather laborious. In addition they use the fact that the Lorentz anomaly is not present in $d=4$ 
 and do not proceed to evaluate the corresponding term.
Hence we  improve on their work in verifying  the vanishing of the Lorentz anomaly explicitly and can do 
so in an economic manner.

\section{The  Fujikawa Method Adapted to Weyl Fermions}
\label{sec:Fujikawa}

The same results can be obtained adapting   the path integral derivation of anomalies by  Fujikawa~\cite{Fujikawa:1979ay,Fujikawa:1980rc} (cf. also \cite{Bertlmann:1996xk,Fujikawa:2004cx}) 
to two-components Weyl fermions. As far as we know, only Dirac fermions with a projector $P_L$ are considered in the literature, which suffer from an ill-defined path integral due to the non-invertibility of $i\sD P_L$.
In the path integral, the anomaly arises
from a non-trivial Jacobian which can be written as  a fraction of determinants  \cite{Filoche:2022dxl}\footnote{Let us comment on the sign  in $\det\left(\dc-\de_\al \dc\right)$. 
In the Fujikawa approach one transforms the path integral variables only and since 
$\de_\alpha S[\tilde\psi,\tilde{\bar\psi},e]=\de_\al\tilde\psi \frac{\de S}{\de \tilde\psi}+\de_\al\tilde{\bar\psi} \frac{\de S}{\de \tilde{\bar\psi}}+\de_\al \tensor{e}{^\mu_a}\frac{\de S}{\de \tensor{e}{^\mu_a}}=0$, this then implies that 
it is $- \de_{\al} \DW$, which stems from $\de_\al \tensor{e}{^\mu_a}\frac{\de S}{\de \tensor{e}{^\mu_a}}$,  that appears under the $\al$-variation.}  
\begin{alignat}{2}
\label{eq:JJ}
& J[\alpha] &\;=\;&  e^{-\int\dd^4x\,e\,\alpha(x)\mathcal{A}(x)}   \nonumber \\[0.1cm]
&  &\;=\;&  \frac{\det( \DW )}{\det( \DW - \de_{\al}  \DW \ )} = \frac{1}{\det( \mathds{1} -  \de_{\al} \DW \, \DW^{-1})}  = \exp(  \Tr[ \de_{\al} \DW \, \DW^{-1}] + \ORD(\al^2))
 \; ,
\end{alignat}
and when expanded assumes the same 
form as in \eqref{eq:Wformal}. This  guarantees that it is well-defined, that is to say the global  phase 
of the determinant cancels  in this expression and the operator  $\de_{\al} \DW \, \DW^{-1}$ maps into the same Hilbert space as mentioned previously.

We proceed to construct the path integral measure.
The Weyl operator \eqref{eq:DW2} is not hermitian and does not have a well-defined eigenvalue problem. However, since $i\sD$ is hermitian and   $i\sD : (\frac{1}{2},0) \oplus  (0,\frac{1}{2}) \to (\frac{1}{2},0) \oplus  (0,\frac{1}{2}) $, 
the eigenvalue problem $i\sD\phi_n=\lambda_n\phi_n$ is well-posed. In particular   
  $\lambda_n\in\mathds{R}$ and the eigenfunctions  $\{\phi_n\}$ define a complete orthonormal basis. These  functions can be decomposed as follows
\begin{equation}
\phi_n=\begin{pmatrix}
\phi^R_n \\
\phi^L_n \\    
\end{pmatrix}\;,
\end{equation}
 where $\{\phi^{R,L}_n\}$  define orthonormal eigenbases of right- and left-handed 
 Weyl fermions respectively . The eigenvalue equations become
\begin{equation}
\DW\phi^L_n=i\bar\sigma^\mu ( \partial + \omega_L)_\mu\phi^L_n=\lambda_n\phi^R_n \;, \qquad 
\DW^\dagger \phi^R_n=i\sigma^\mu ( \partial + \omega_R)_\mu \phi^R_n=\lambda_n\phi^L_n\;.
\label{eigevanlueEqLR}
\end{equation}
We thus decompose the path integral measure into these eigenbases
\begin{equation}
\label{eq:eigen}
\tilde{\psi}_L=\sqrt{e}\sum_n a_n \phi^L_n \;, \qquad 
\tilde{\bar\psi}_L=\sqrt{e}\sum_n \bar{b}_n\left(\phi^R_n\right)^\dagger\; ,
\end{equation}
for which the measure assumes the form\footnote{This change of variable is defined up to a phase. As emphasised in \SEC\ref{sec:detD}, this phase is irrelevant when dealing with the covariant form of anomalies which is the case here~\cite{Fujikawa:2004cx}.}
\begin{equation}
 \DM\tilde\psi_L \DM\tilde{\bar{\psi}}_L = \prod_n \dd a_n\dd \bar b_n\;.
\end{equation}
{In order to evaluate the Jacobian, we need to determine how a variation $\de_{\al}$ acts on the Weyl fermion 
 $\tilde\psi'_L \equiv \tilde\psi_L+\de_\al \tilde\psi_L$ and its barred counterpart.  
 We expand $\tilde\psi'_L$ and $\tilde{\bar\psi}'_L$ into the
eigenbases as in \eqref{eq:eigen}} to obtain
\begin{equation}
a_m'=\sum_n\left(\delta_{mn}+A_{mn}\right)a_n \;, \qquad 
\bar b_m'=\sum_n\left(\delta_{mn}+B_{mn}\right)\bar b_n\; .
\end{equation}
where $A$ and $B$ are of $\ORD(\al)$, 
and the resulting Jacobian of the Grassmann variables reads
\begin{equation}
\log J=-\Tr\,A-\Tr\,B+\mathcal{O}(\alpha^2)\;.
\end{equation}
Using the orthonormality of the eigenbasis, we obtain the Jacobians of the transformations $\delta^W_\sigma$, $\delta^d_\xi$ and $\delta^L_\alpha$,
\begin{alignat}{2}
\label{eq:lnFuj}
&\log J_{\mathrm{\WEYL}}[\sigma] &\;=\;& -\int\dd^4x\,e\,\frac{\sigma}{2}\sum_n\left((\phi^R_n)^\dagger\phi^R_n+  \{ R \leftrightarrow L\} \right) \;,
\nonumber  \\[0.1cm]
&\log J_{\mathrm{diffeo}}[\xi^\mu] 
&\;=\;& - \int\dd^4x\,e\,\sum_n\Big((\phi^R_n)^\dagger(\xi^\mu \nabla_\mu+\frac{1}{2}(D_\mu\xi^\mu))\phi^R_n 
- \{ R \leftrightarrow L\} \Big) \;, \nonumber  \\[0.1cm]
&\log J_{\mathrm{Lorentz}}[\alpha_{ab}]
&\;=\;& -\int\dd^4x\,e\,\frac{1}{2}\alpha_{ab}\sum_n\Big((\phi^R_n)^\dagger\lambda^{ab}\phi^R_n - \{ R,\la \leftrightarrow L,\mu \}    \Big)\;.
\end{alignat}
These expression are ultraviolet divergent. They can be  regularised 
by introducing $e^{-\frac{\lambda_n^2}{\Lam^2}}$ into the sums,  following Fujikawa. 
Note that $J_{\mathrm{diffeo}}$ and $J_{\mathrm{Lorentz}}$ are fully determined by the zero modes  whereas $J_{\mathrm{trace}}$ is determined by the zero modes only at lowest order
of the loop expansion \cite{Fujikawa:2004cx}. This means that the diffeo and Lorentz anomalies 
are of topological nature and determined at the one-loop level, as also argued by 
 \cite{Alvarez-Gaume:1983ihn}.

Using the eigenvalue equation Eq.~\eqref{eigevanlueEqLR} the following replacement 
can be made 
\begin{equation}
e^{-\frac{\lambda_n^2}{\Lam^2}}\phi^L_n = e^{-\frac{\DW^\dagger\DW}{\Lam^2}}\phi^L_n\; ,\qquad e^{-\frac{\lambda_n^2}{\Lam^2}}\phi^R_n = e^{-\frac{\DW\DW^\dagger}{\Lam^2}}\phi^R_n\;,
\end{equation}
where $\DW$ is the Weyl operator \eqref{eq:DW2}. 
As in the previous section
the Jacobians can be recast in Dirac space using Eq.~\eqref{WeylToDirac} 
\begin{alignat}{3}
\label{JacobiansFinal}
&\log J_{\mathrm{\WEYL}}[\sigma] &\;=\;&-\de^W_\sigma W &\;=\;& -\frac{1}{2}\lim_{\Lam\to\infty}\Tr\,\sigma e^{-\frac{(i\sD)^2}{\Lam^2}}  \;, \nonumber \\[0.1cm]
&\log J_{\mathrm{diffeo}}[\xi] &\;=\;&-\de^d_\xi W&\;=\;& + \frac{1}{2}\lim_{\Lam\to\infty}\Tr\,\gamma_5 \left(\xi^\mu \nabla_\mu+\frac{1}{2}(
D_\mu\xi^\mu)\right) e^{-\frac{(i\sD)^2}{\Lam^2}}=0 \;, \nonumber \\[0.1cm]
&\log J_{\mathrm{Lorentz}}[\alpha] &\;=\;&-\de^L_\al W&\;=\;& - \frac{1}{2}\lim_{\Lam\to\infty}\Tr\,\alpha_{ab}\Sigma^{ab}\gamma_5 e^{-\frac{(i\sD)^2}{\Lam^2}}=0 \;.
\end{alignat}
These results are identical to Eqs.~\eqref{deltad_Weyl}, \eqref{deltaL_Weyl} and \eqref{eq:WW} found 
in the proper time regularisation and can thus  be seen as a confirmation.

\section{Comments on the Literature}
\label{sec:comments}

In view of the variety of methods applied  and results obtained in the literature we make 
an attempt at understanding the results.

\subsection{Approaches finding a non-vanishing $\RRt$-term }

Let us first  turn to the approaches finding  $ e  \neq 0$ in \EQ\eqref{Eq1Weyl} which are in disagreement with our result. 
\begin{itemize}
\item The perturbative approaches require an expansion in the  metric $g_{\al \be}  = \eta_{\al \be} + h_{\al \be}$ followed by a ``covariantisation" procedure, towards the end of the computation, to go beyond the linearisation.
In practice this involves computing Feynman diagrams with a spectator Weyl fermion, that is {$\Lagr=e\,\bar\psi\left(i\sD P_L + i\slashed\pa P_R\right)\psi$}, such that a propagator can be obtained by inversion. 
In a gauge theory  this procedure has been successfully applied \cite{Alvarez-Gaume:1983ict,Alvarez-Gaume:1984zlq} but in the context of gravity this is more subtle 
since the right-handed spectator seems to violate the universality of gravity (Lorentz invariance).\footnote{{ In a gauge theory with a right-handed spectator, the gauge symmetry is preserved by adapting the gauge transformation  $ e^{i\al}\psi\to  e^{i\al P_L}\psi$. However, the Lorentz transformation cannot be adapted in this manner since 
the transformation of the vierbein in $\slashed\pa P_R \psi =\gamma^a\tensor{e}{^\mu_a}\pa_\mu P_R  \psi$ would 
spoil  Lorentz invariance. 
In other words the spectator ceases to be a spectator in the context of gravity.}}  
Between the perturbative approaches the main open question is the choice of regularisation procedure. 
In  \cite{Bonora:2014qla,Bonora:2015nqa,Bonora:2017gzz,Bonora:2018ajz,Bonora:2018obr,Bonora:2022izj,Liu:2022jxz} 
dimensional, differential or Pauli-Villars regularisation have been employed, while verifying the vanishing of the diffeo anomaly, and a non-zero $\RRt$-term has 
been found.  
\item The Fujikawa method was applied in \cite{Liu:2022jxz} relying on a mass for a left-handed field $\bar\psi m P_L\psi$ which is known to vanish. In~\cite{Liu:2023yhh}, a regularisation with  a mass term $\bar\psi m^A P_L\psi$ such that $\{m^A,\gamma_5\}=[m^A,\gamma^\mu]=0$ and $(m^A)^\dagger=m^A$ was introduced. 
We find that the only matrix satisfying these equations is $m^A=0$ and hence 
the method  seems contradictory.  
\item The heat kernel approaches were applied in \cite{Bonora:2014qla,Bonora:2018obr}.
\begin{itemize}
\item In \cite{Bonora:2014qla} the heat kernel regulator 
$\exp{(-{\DW^\dagger \DW}/{\Lam^2})}$ was used where $\DW$ is the Weyl operator of 
$\psi_L$. 
This differs from \eqref{eq:WW}, see also \eqref{eq:lnFuj} (as well as  \EQ 4.4 in \cite{Leutwyler:1985ar}), which corresponds to $\exp{(-{\DW^\dagger \DW}/{\Lam^2})} + \exp{(-{\DW \DW^\dagger}/{\Lam^2})}$ thereby  explaining the difference.
{In our formalism the sum is a direct consequence of the variation \eqref{eq:var} 
and the algebraic identity \eqref{eq:crucial}.}\footnote{{One may be tempted to think that regularising with $\exp{(-{\DW^\dagger \DW}/{\Lam^2})} $ only would be correct since it maps left- onto left-handed fermions.  
As emphasised earlier the  $\DW$-operator maps from one chirality into the other 
and this indicates that  both $\exp{(-{\DW^\dagger \DW}/{\Lam^2})}$ and $\exp{(-{\DW \DW^\dagger}/{\Lam^2})}$ ought to appear in the regularisation.}} 
\item  In \cite{Bonora:2018obr} the metric-axial gravity (MAT) formalism was applied, where 
 the metric is supplemented  by  an axial part  
$g_{\mu \nu} \to g_{\mu \nu} + \ga_5 f_{\mu\nu}$, motivated 
by earlier work by Bardeen \cite{Bardeen:1969md}.
However, concerning the regularisation the same remarks apply as above.
\end{itemize}
\end{itemize}

\subsection{Approaches finding a vanishing $\RRt$-term }

Let us  discuss the approaches agreeing with our result $e = 0$ in \EQ\eqref{Eq1Weyl}.
\begin{itemize}
\item The work of Leutwyler and Mallik \cite{Leutwyler:1985ar} stands out in that it seems to be unknown to the current community. 
They invented the method \eqref{eq:lnreg} which we use throughout. They use point splitting and the proper time regularisation
and evaluate the expression in terms of the heat kernel. 
We  agree with their findings. As mentioned previously the difference in our paper is that we use the CDE {which is more straightforward, and we also verify the vanishing of the Lorentz anomaly explicitly}. 
We  further adapt  Fujikawa's method for Weyl fermions. 
\item In \cite{Bastianelli:2016nuf}  Fujikawa's method was applied 
writing a Majorana type mass term for the Weyl fermion
which then regulates the determinant. 
This is possible since the Weyl representation in Euclidian space is real in $d=4$ \cite{Alvarez-Gaume:1983ihn}.   
The problem of the determinant of the Weyl operator, which we have addressed in detail, has not been considered 
explicitly in that paper.\footnote{In \cite{Bonora:2017gzz} doubts were expressed 
that introducing a Majorana  mass is legitimate.  It remains however unclear to us why the Majorana mass term 
as first proposed in \cite{Alvarez-Gaume:1983ihn} should not be legitimate}.
\item  The MAT approach   was applied with the Fujikawa method
\cite{Bastianelli:2019zrq} again using a  Majorana-type mass.  
The definition of the Weyl operator has not been addressed either.
\item A non-perturbative approach employing the Hadamard regularisation was carried out in \cite{Frob:2019dgf} working directly in the EMT. For the Weyl fermion the EMT is taken to be symmetric hence the vanishing of the Lorentz anomaly is assumed and not verified.
\item In another series of perturbative approaches dimensional regularisation \cite{Abdallah:2021eii} (along with the Breitenlohner-Maison $\gamma_5$-scheme) have been applied. 
These computations are further reviewed and clarified in \cite{Abdallah:2023cdw}. 
Diffeomorphism invariance is verified assuming that there is no Lorentz anomaly. 
\end{itemize}

The main mystery remains why some perturbative approaches obtain $e \neq 0$ and others $e =0$, even though the vanishing of the diffeo anomaly at lowest order of the metric expansion is verified in each case. 
It either means that there is an error in one of the 
computations, which tend to be lengthy, or that there is a specific problem with 
some of the regularisations.   
{As for the non-perturbative approaches, only Leutwyler and Mallik verify the vanishing of the diffeo anomaly in \cite{LEUTWYLER198565,Leutwyler:1985ar} (but not the Lorentz anomaly), whereas the other papers 
do not address them at all.}

\section{Dimensions Other than $d=4$}
\label{sec:d}

In $d=4$ we concluded that for free Weyl fermions  the $\RRt$-term is absent in the trace anomaly.
It is therefore natural to ask whether  $P$- and or $CP$-odd terms could be present in any other even 
dimension.\footnote{{Odd dimensional spaces are beyond the scope of this paper and we refer the reader 
to \cite{Alvarez-Gaume:1984zst} for intricate relations with the even dimensional case.}}
In fact up to
Eqs.~\eqref{deltad_Weyl}, \eqref{deltaL_Weyl} and \eqref{eq:WW} the   expressions
 are independent of the dimension in our approach.  
 In particular the factor $\frac{1}{2}$ in \eqref{eq:WW} makes it clear that 
 the trace anomaly of a Weyl fermion 
remains half the trace anomaly of a Dirac fermion in any even 
dimensional spacetime.
Even in the absence of concrete computations one can make interesting observations 
and come to the same conclusion:
\begin{itemize} 
\item It is important to distinguish $d = 2\;(\textrm{mod } 4)$ and 
$d = 4\;(\textrm{mod } 4)$  as their (Euclidean)  Weyl representation 
are complex  and real respectively  \cite{Alvarez-Gaume:1983ihn}. For example,  
Lorentz and diffeo anomalies cannot   
arise in $d = 4\;(\textrm{mod } 4)$ dimensions since the reality of the representation allows for 
a Pauli-Villars  regulator mass term  which is symmetry-preserving and this means that no 
diffeo and Lorentz anomaly can appear  \cite{Alvarez-Gaume:1983ihn}. 
Another way to look at it is that  
 in $d = 4\;(\textrm{mod } 4)$, unlike in $d = 2\;(\textrm{mod } 4)$, 
the $CPT$ operation flips the chirality such that the Weyl fermions effectively look
 vector-like  \cite{Alvarez-Gaume:1983ihn} (and \cite{Alvarez-Gaume:1985zzv} for more detail).

\item In $d = 2\;(\textrm{mod } 4)$ a $P$- and $CP$-odd term should not  violate $CPT$ since the appearance of the  factor of $i$ is dimension-dependent. 
For example in Minkowski space one has 
 $\Tr[\ga_\al \ga_\be \ga_5 ] = 2 \eps_{\al\be}$ in $d=2$ 
whereas  $\Tr[\ga_\al \ga_\be \ga_\ga \ga_\de \ga_5 ] = 4i \eps_{\al\be\ga\de}$ in $d=4$,
and this is related to 
the complex  and real  representations mentioned above.  Hence  in $d = 2\;(\textrm{mod } 4)$   
we would not expect an imaginary prefactor and thus no $CPT$-violation.
\item  However, one can argue that in $d = 2\;(\textrm{mod } 4)$ one cannot write down 
a $P$- or $CP$-odd {diffeo-invariant scalar} of mass dimension $d$ (which is relevant for {the trace} anomaly).  Firstly, let us note that due to the  Bianchi identity, $\epsilon^{\dots \mu\nu\rho}R_{\alpha\mu\nu\rho}=0$ holds in any dimension, that is to say the Levi-Civita tensor cannot contract more than two indices of a Riemann tensor.
For concreteness let us first focus on $d=6$. Using the Bianchi identities one can show that 
a parity-odd operator of mass dimension 6 can only be of the form,
\begin{equation}
\epsilon^{\alpha_1\dots \alpha_6}R_{\alpha_1\alpha_2 . .}R_{\alpha_3\alpha_4 . .}R_{\alpha_5\alpha_6 . .}\; .
\end{equation}
The only way to contract these Riemann tensors together is
\begin{equation}
\epsilon^{\alpha_1\dots \alpha_6}\tensor{R}{_{\alpha_1}_{\alpha_2}_\mu^\nu}\tensor{R}{_{\alpha_3}_{\alpha_4}_\rho_\nu}\tensor{R}{_{\alpha_5}_{\alpha_6}^\rho^\mu}=0\;,
\end{equation}
which vanishes since $\epsilon^{\alpha_1\dots \alpha_6}\tensor{R}{_{\alpha_1}_{\alpha_2}_\mu^\nu}\tensor{R}{_{\alpha_3}_{\alpha_4}_\rho_\nu}$ is symmetric under $\mu\leftrightarrow\rho$ whereas $\tensor{R}{_{\alpha_5}_{\alpha_6}^\rho^\mu}$ is antisymmetric. On the other hand, in $d=8$ for example, there is an even number of Riemann tensors, which can then be contracted in pairs
\begin{equation}
\epsilon^{\alpha_1\dots \alpha_8}\tensor{R}{_{\alpha_1}_{\alpha_2}_\mu_\nu}\tensor{R}{_{\alpha_3}_{\alpha_4}^\mu^\nu}\tensor{R}{_{\alpha_5}_{\alpha_6}_\rho_\sigma}\tensor{R}{_{\alpha_7}_{\alpha_8}^\rho^\sigma}\neq0\, .
\end{equation}
This  generalises straightforwardly   to any even dimension
\begin{equation}
\epsilon^{\alpha_1\dots \alpha_{2n}}\tensor{R}{_{\alpha_1}_{\alpha_2}_._.}\dots\tensor{R}{_{\alpha_{2n-1}}_{\alpha_{2n}}_._.}  \left\{\begin{array}{ll} = 0  & d = 2\;(\textrm{mod } 4) \\
\neq 0  & d = 4\;(\textrm{mod } 4) 
     \end{array} \right.\;,
\end{equation}
since it involves an odd number of Riemann tensors in $d = 2\;(\textrm{mod } 4)$ but an even number in $d = 4\;(\textrm{mod } 4)$.
In fact, the absence of such a term in $d=6$ has been inferred from a cohomology-type  argument given a long time ago \cite{Bonora:1985cq}.
\end{itemize}
In summary the absence of parity-odd terms can be established in $d = 2\;(\textrm{mod } 4)$ without explicit computation whereas in $d = 4\;(\textrm{mod } 4)$ a computation is required. 

\section{Conclusion} 
\label{sec:conc}

We investigated the trace anomaly of a free Weyl fermion in a curved space which gave rise to 
some controversy recently as some authors found a term proportional to the Pontryagin density $\RRt$, as reviewed in  \cite{Bonora:2022izj}.  
The results found  in the literature for $d=4$, as we point out, do not only violate 
$P$ and $CP$ but also $CPT$ because of the purely imaginary prefactor.  
Handling Weyl fermions is technically subtle and requires care as the Weyl
determinant is ill-defined.
{We use the method of proper time regularisation building on earlier work by  Leutwyler 
(\SEC\ref{sec:proper}) and develop the Fujikawa method for a two-component Weyl fermion 
(\SEC\ref{sec:Fujikawa}), where it is not the  Weyl determinant but its well-defined  variation which is used. 
 This allows us  to perform the computation in a rather compact manner from which our main result follows:
  the 
 trace anomaly of the Weyl fermion is half the one of a Dirac fermion and therefore no Pontryagin density 
 $\RRt$ is present.}
We have further checked that the diffeomorphism  and the Lorentz anomalies are absent 
which is a consistency check  on our regularisation since these anomalies are known to be absent in $d = 4\;(\textrm{mod } 4)$  \cite{Alvarez-Gaume:1983ihn}.
In \SEC\ref{sec:d} we have concluded that no parity-odd term can appear in any even dimension  for a free
Weyl fermion. 
In \SEC\ref{sec:comments} we have attempted to compare to the literature 
and thereby hope that our findings are helpful in settling the controversy.  
An investigation of the anomalies from a completely different viewpoint is  presented in \cite{Larue:2023qxw}.

Our findings do not mean that a $\RRt$-term could not play a role in fundamental physics. It can arise from  sources other than Weyl fermions as studied in extensions of gravity  \cite{Alexander:2009tp}, it may appear in conjunction with axions  \cite{Dvali:2013cpa,Alonso:2017avz}, more generally in connection with $P$- and  $CP$-violation e.g. \cite{Deser:1980kc} 
and we note that it has been debated in the context of conformal field theories \cite{Coriano:2023cvf}.
 Further exploration is left to future work.

\section*{Acknowledgments} 
The authors are grateful to Luis \'Alvarez-Gaum\'e, Bill Bardeen, Fiorenzo Bastianelli, Loriano Bonora, Mike Duff, Sebastián Franchino-Viñas for helpful discussions or comments on the manuscript. The authors are particularly thankful to Sebastián Franchino-Viñas for rising their curiosity on the subject.
The work of RL and JQ is supported by the IN2P3 Master projects A2I and BSMGA, by the project AFFIRM of the Programme National GRAM of CNRS/INSU with INP and IN2P3 co-funded by CNES and by the project EFFORT supported by the programme IRGA from UGA. JQ thanks LAPTh Annecy for hospitality while this work was completed. JQ and RZ acknowledge the  support of CERN associateships. The work of RZ is supported by the  STFC Consolidated Grant, ST/P0000630/1. 
Many manipulations were carried out with the help of Mathematica and the package xAct \cite{xAct}.


\appendix

\section{Weyl Fermions in Curved Spacetime}
\label{App:WeylFermion}

The Dirac matrices can be expressed as
\begin{equation}
\gamma^a=\begin{pmatrix}0 & \bar\sigma^a\\
\sigma^a &0\end{pmatrix},\quad\quad \gamma_5=\begin{pmatrix} \mathds{1}_2 & 0 \\ 0 & -\mathds{1}_2\end{pmatrix}\;,
\end{equation}
where $\sigma^a=(\mathds{1}_2,\sigma^i)$, $\bar\sigma^a=(\mathds{1}_2,-\sigma^i)$, and $\sigma^{i=1,2,3}$ are the Pauli matrices. As for the Dirac matrices, we have $\sigma^\mu=\tensor{e}{^\mu_a}\sigma^a$.
In an Euclidian space, it is possible to choose a representation of the $\sigma$-matrices such that $\bar\sigma^\mu=(\sigma^\mu)^\dagger$, i.e $(\gamma^\mu)^\dagger=\gamma^\mu$ which is the convention used throughout.
From $\{\gamma^\mu,\gamma^\nu\}=2g^{\mu\nu}\mathds{1}_4$ one deduces 
\begin{align}
\sigma^\mu\bar\sigma^\nu+\sigma^\nu\bar\sigma^\mu=\bar\sigma^\mu\sigma^\nu+\bar\sigma^\nu\sigma^\mu=2 g^{\mu\nu}\mathds{1}_2\;.
\end{align} 
The generator of rotations can be written as 
\begin{equation}
\Sigma^{ab}=\frac{1}{4}[\gamma^a,\gamma^b]=\begin{pmatrix}
\lambda^{ab} & 0 \\ 0 & \mu^{ab}    
\end{pmatrix} \;,
\end{equation}
where
\begin{equation}
\lambda^{ab}=\frac{1}{4}(\bar\sigma^a\sigma^b-\bar\sigma^b\sigma^a),\quad\quad \mu^{ab}=\frac{1}{4}(\sigma^a\bar\sigma^b-\sigma^b\bar\sigma^a) \;.
\end{equation}
Using $D_\mu\psi=(\pa_\mu+\frac{1}{2}\omega_{\mu,ab}\Sigma^{ab})\psi$ and decomposing $\psi$ and $D_\mu$ as\footnote{$D_\mu$ is diagonal since it does 
not change the chirality of fermions.}
\begin{equation}
\psi=\begin{pmatrix}
    \psi_R \\ \psi_L
\end{pmatrix},\quad\quad D_\mu=\begin{pmatrix}
    D^R_\mu & 0 \\ 0 & D^L_\mu
\end{pmatrix}\;,
\end{equation}\\
one obtains
\begin{alignat}{3}
\label{eq:DLR}
&D^R_\mu\psi_R&\;=\;&(\pa_\mu+\omega^R_\mu)\psi_R&\;=\;&(\pa_\mu+\frac{1}{2}\omega_{\mu,ab}\lambda^{ab})\psi_R \;, \nonumber \\[0.1cm]
&D^L_\mu\psi_L&\;=\;&(\pa_\mu+\omega^L_\mu)\psi_L&\;=\;&(\pa_\mu+\frac{1}{2}\omega_{\mu,ab}\mu^{ab})\psi_L\;.
\end{alignat}
The Weyl operator, used in the main text,  can be identified from the second line.
Note that when acting on a scalar with Lorentz index such as $\xi^\mu$ one has $(D^R_\mu\xi^\nu)=(D^L_\mu\xi^\nu)=(D_\mu\xi^\nu)=(\nabla_\mu\xi^\nu)=(\pa_\mu\xi^\nu)+\Gamma^\nu_{\mu\rho}\xi^\rho$.

From the compatibility with Dirac matrices $[D_\mu,\gamma^\nu]=0$ one obtains
\begin{align}
\label{eq:DRsigma}
D^R_\mu\bar\sigma^\nu=\bar\sigma^\nu D^L_\mu,\quad\quad D^L_\mu\sigma^\nu=\sigma^\nu D^R_\mu\;.
\end{align}
Finally, since
\begin{equation}\label{eq:iDtoLR}
i\sD=\begin{pmatrix}
0 & i\bar\sigma^\mu D^L_\mu \\
i\sigma^\mu D^R_\mu & 0
\end{pmatrix}\;,
\end{equation}
is hermitian in Euclidian space, it follows that 
\begin{equation}
\left(i\bar\sigma^\mu D^L_\mu\right)^\dagger=i\sigma^\mu D^R_\mu\;.
\end{equation}

\section{Covariant Derivative Expansion  -- Computations}
\label{App:CDEcomputations}

In this Appendix, we outline the computation of the anomalies of a Weyl fermion using the CDE in curved spacetime \cite{Larue:2023uyv}.

Let us first note that we regularised the functional traces using the function $e^{-x}$ in both sections. The result is however independent of this choice. In fact, any smooth function $f$ such that $f(0)=1$ and $x^n f^{(n)}(x)\to 0$ for $x\to\infty$ and for all  $n\geq0$ can be used instead. This is well-known for  Fujikawa's approach \cite{Fujikawa:2004cx} 
and we establish it here for 
 Leutwyler's approach. For a function $f$ satisfying the criteria above the following equation
 holds
\begin{equation}
\dc^{-1}=\dc^\dagger\int_0^\infty\dd t\,  f'(t\dc\dc^\dagger)\;,
\end{equation}
for which \eqref{eq:proper} is a special case.

Having established this universality in the function $f$  we turn to 
the CDE for which it is convenient to use $f(x)=1/(1+x)$. We thus consider the following functional traces\footnote{Note that if we had chosen a representation of the Dirac matrices such that $(\gamma^\mu)^\dagger=-\gamma^\mu$ in Euclidian space, we would obtain $f\left(\frac{\sD^2}{\Lam^2}\right)$ instead of $f\left(\frac{(i\sD)^2}{\Lam^2}\right)=f\left(\frac{\sD^2}{-\Lam^2}\right)$. The anomaly being of order $\Lam^0$ does not depend on that choice.}
\begin{alignat}{3}
\label{TracesGeneric}
&T_1[a]&\;=\;&\lim_{\Lam\to\infty}\Tr\left[a(x) f\left( \frac{(i\sD)^2}{\Lam^2} \right)\right]&\;=\;&\lim_{\Lam\to\infty}\Tr\left[a(x)\frac{\Lambda^2}{-\sD^2+\Lambda^2}\right]\; ,\\[0,1cm]
&T_2[b]&\;=\;&\lim_{\Lam\to\infty}\Tr\left[b^\mu(x) \nabla_\mu\, f\left( \frac{(i\sD)^2}{\Lam^2} \right)\right]&\;=\;&\lim_{\Lam\to\infty}\Tr\left[b^\mu(x) \nabla_\mu\, \frac{\Lambda^2}{-\sD^2+\Lambda^2}\right]\;,\nonumber
\end{alignat}
where $a(x)$ and $b^\mu(x)$ are local functions. 
As in the main text it is understood that the $\Lam^2$- and $\Lam^4$-divergences are subtracted.  

\subsection{Computation of the trace and Lorentz anomalies}

From $T_1$ the trace and the Lorentz anomalies of a Weyl fermion follow
\begin{equation}
\label{eq:traceLorentz}
\delta^W_\sigma W = \frac{1}{2}T_1\left[\sigma\right]\;, \quad 
\delta^L_\alpha W = \frac{1}{2}T_1\left[\alpha_{ab}\Sigma^{ab}\gamma_5\right]\; .\end{equation}
The functional trace can be recast in momentum space as
\begin{alignat}{2}
&T_1[a]&\;=\;&\lim_{\Lam\to\infty}\int\dd^4x\frac{\dd^4q}{(2\pi)^4}\tr\,a(x)\Lambda^2\frac{1}{-(\sD+i\slashed q)^2+\Lambda^2}\; \\[0,1cm]
& &\;=\;&-\lim_{\Lam\to\infty}\int\dd^4x\dfrac{\dd^4q}{(2\pi)^4}\tr\,a(x)(-\Lambda^2)\sum_{n\geq0}\Big[\Delta\big(D^2+g^{\mu\nu}\{D_\mu,iq_\nu\}+\Sigma\cdot F\big)\Big]^n\Delta\;,\nonumber
\end{alignat}
where $\Delta=1/(q^2+\Lambda^2)$, $\Sigma^{\mu\nu}=[\gamma^\mu,\gamma^\nu]/4$, and $\tr$ denotes the trace in internal space. Note that we can maintain $\pa_\mu q_\nu =0$, but contrary to the CDE in flat spacetime we have $D_\mu q_\nu=-\Gamma^\rho_{\mu\nu}q_{\rho}\neq0$ and $[D_\mu,\Delta]\neq0$~\cite{Larue:2023uyv}. 
$F$ is the fermion field strength due to the spin-connection
such that for any fermion $\tilde\psi$ we have 
$[D_\mu,D_\nu]\tilde\psi=F_{\mu\nu}\tilde\psi$ with\footnote{Note that $\psi$ and $\tilde\psi$ have the same field strength.}
\begin{equation}
\label{eq:FsigmaF}
F_{\mu\nu}\;=\;\frac{1}{4}\gamma^\rho\gamma^\sigma R_{\mu\nu\rho\sigma}\; ,\quad\quad \Sigma\cdot F\;=\;-\frac{R}{4}\mathds{1}_{\mathrm{Dirac}}\; .
\end{equation}
The expansion is then carried out with the help of Mathematica and the package xAct \cite{xAct}. The result reads
\begin{align}
\label{eq:T1b}
T_1[a]=\frac{1}{16\pi^2}&\int\dd^4x\, \tr\,a(x) \Bigg\{-\frac{1}{6}\Box\left(\Sigma\cdot F\right)-\frac{1}{12}F^2-\frac{1}{72}R^2+\frac{1}{180}R_{\mu\nu}R^{\mu\nu}\; \\[-,1cm]
&-\frac{1}{180}R_{\mu\nu\rho\sigma}R^{\mu\nu\rho\sigma}-\frac{1}{30}\Box R-\frac{1}{6}R\,\Sigma\cdot F-\frac{1}{2}\left(\Sigma\cdot F\right)^2\Bigg\}\; ,\nonumber
\end{align}
from which, by using \eqref{eq:traceLorentz} and \eqref{eq:FsigmaF}, the trace anomaly 
and the vanishing of the Lorentz anomaly follows.

In fact once we know, from \eqref{eq:T1b}, that the Lorentz anomaly is covariant we can infer its vanishing in 
yet another way in $d=4$. 
It can be shown, using intrinsically 4-dimensional identities~\cite{Remiddi:2013joa,Chala:2021cgt,Chung:2022ees}, that the only parity-odd (covariant) 2-tensor of mass dimension 4 is $g^{\mu\nu}R\tilde R$, which is symmetric in its indices and thus vanishes when contracted with the Lorentz parameter $\alpha_{\mu\nu}$.
In particular, one has in $d=4$
\begin{equation}
\frac{1}{2}g^{\alpha\beta}\RRt = \epsilon^{\alpha\nu\rho\sigma}\tensor{R}{^\beta_\nu^\lambda^\chi}\tensor{R}{_\rho_\sigma_\lambda_\chi}=\epsilon^{\mu\nu\rho\sigma}\tensor{R}{^\alpha^\lambda_\mu_\nu}\tensor{R}{^\beta_\lambda_\rho_\sigma}\; ,
\end{equation}
and
\begin{equation}
\epsilon^{\alpha\nu\rho\sigma}\tensor{R}{^\beta^\lambda_\rho_\sigma}\tensor{R}{_\nu_\lambda}=0\; ,
\end{equation}
for the Ricci-tensor contraction.
Every other parity-odd 2-tensor of mass dimension 4 is related to these by Bianchi identities and algebra.

\subsection{Computation of the diffeomorphism anomaly}

The diffeo anomaly is given by
\begin{equation}
\label{eq:computediffeo1}
\delta^d_\xi W=-T_2[\xi\gamma_5]-\frac{1}{2}T_1[(D_\mu\xi^\mu)\gamma_5]\; ,
\end{equation}
a sum of a $T_1$- and a $T_2$-term \eqref{TracesGeneric}. 
Let us focus on the $T_1$-term first.
Using \eqref{eq:FsigmaF}  the only term that is non-vanishing under the Dirac trace in $T_1[(D_\mu\xi^\mu)\gamma_5]$ is
\begin{alignat}{2}
\label{eq:computediffeo2}
&T_1[(D_\mu\xi^\mu)\gamma_5]&\;=\;&\frac{-1}{16\pi^2}\int\dd^4x\,e\,(D_\mu \xi^\mu)\,\tr\,\gamma_5\,\left(-\frac{1}{12}F^2\right) \\[0,1cm]
& &\;=\;&\frac{-1}{16\pi^2}\int\dd^4x\,e\, \xi^\mu\,\tr\,\gamma_5\,\left(\frac{1}{12}[D_\mu,F^2]\right)\;,\nonumber
\end{alignat}
where integration by parts was applied using the fact that  $\xi^\mu(x)$ vanishes at infinity.

Finally, let us turn to the 
 $T_2$-term which is less straightforward. 
In \EQ\eqref{TracesGeneric}, it is convenient to rewrite $\nabla=D-\omega$, such that
\begin{equation}
T_2[b]=\lim_{\Lam\to\infty}\Tr\left[b^\mu(x) D_\mu\, \frac{\Lambda^2}{-\sD^2+\Lambda^2}\right]-\lim_{\Lam\to\infty}\Tr\left[b^\mu(x) \omega_\mu\, \frac{\Lambda^2}{-\sD^2+\Lambda^2}\right]\;.
\end{equation}
We can notice that the second term is the Lorentz anomaly with $\alpha_{ab}=b^\mu \omega_{\mu ab}$ (with $b^\mu\propto\gamma_5$), hence vanishes as we just verified. We are left with
\begin{alignat}{2}
&T_2[b]&\;=\;&\lim_{\Lam\to\infty}\Tr\left[b^\mu(x) D_\mu\, \frac{\Lambda^2}{-\sD^2+\Lambda^2}\right] \\[0,1cm]
& &\;=\;&\lim_{\Lam\to\infty}\int\dd^4x\frac{\dd^4q}{(2\pi)^4}\tr\,b^\mu(x)(D_\mu+iq_\mu)(-\Lambda^2)\sum_{n\geq0}\Big[\Delta\big(D^2+g^{\mu\nu}\{D_\mu,iq_\nu\}+\Sigma\cdot F\big)\Big]^n\Delta\; .\nonumber
\end{alignat}
The computation must not be carried out in a manifestly covariant manner, and one cannot use a specific choice of coordinate (for example Riemann normal coordinates). 
Since the diffeo anomaly involves $b^\mu(x)=\xi^\mu(x)\gamma_5$, the computation can be simplified using $\Tr \,\gamma_5=\Tr \,\gamma^\mu\gamma^\nu\gamma_5=0$, 
and noticing that the only source of Dirac matrices are in the covariant derivatives via the spin-connection.
We finally obtain a very compact result
\begin{equation}
\label{eq:computediffeo3}
T_2[\xi\gamma_5]=\frac{-1}{16\pi^2} \int\dd^4x\,e\,\xi^\mu\,\tr\,\gamma_5  \left(-\frac{1}{24}[D_\mu,F^2] \right)\;,
\end{equation}
as every other term vanishes by lack of Dirac matrices.
 
Using \eqref{eq:computediffeo1}, \eqref{eq:computediffeo2} and \eqref{eq:computediffeo3} we see the canceling of terms and finally conclude 
 that the diffeo anomaly of a Weyl fermion vanishes.

\bibliographystyle{JHEP}
\bibliography{biblio}

\end{document}